\documentclass[a4paper]{article}

\usepackage{INTERSPEECH2020}
\usepackage{mathtools}

\usepackage{amsmath, lipsum}
\usepackage{amssymb}
\DeclareMathOperator*{\argmax}{arg\,max}

\usepackage{multicol}
\usepackage{xcolor}
\usepackage[title]{appendix}
\usepackage{lipsum}
\usepackage{enumitem}

\newcommand*{\rom}[1]{\expandafter\@slowromancap\romannumeral #1@}
\newcommand{\RNum}[1]{\uppercase\expandafter{\romannumeral #1\relax}}

\usepackage{stfloats}

\title{Do End-to-End Speech Recognition Models Care About Context?}

\name{Lasse Borgholt$^{1, 2}$, Jakob D. Havtorn$^2$, {\v Z}eljko Agi{\' c}$^2$, Anders Søgaard$^1$, Lars Maaløe$^2$, Christian Igel$^1$}

\address{
  $^1$Department of Computer Science, University of Copenhagen, Denmark\\
  $^2$Corti, Copenhagen, Denmark}

\email{\{borgholt, sogaard, igel\}@di.ku.dk, \{lb, jdh, za, lm\}@corti.ai}

\begin{document}

\maketitle
\begin{abstract}
    The two most common paradigms for end-to-end speech recognition are connectionist temporal classification (CTC) and attention-based encoder-decoder (AED) models. It has been argued that the latter is better suited for learning an implicit language model. We test this hypothesis by measuring temporal context sensitivity and evaluate how the models perform when we constrain the amount of contextual information in the audio input. We find that the AED model is indeed more context sensitive, but that the gap can be closed by adding self-attention to the CTC model. Furthermore, the two models perform similarly when contextual information is constrained. Finally, in contrast to previous research, our results show that the CTC model is highly competitive on WSJ and LibriSpeech without the help of an external language model.
\end{abstract}

\section{Introduction}


%
%
\begin{figure*}[b]
  \centering
  \includegraphics[width=\linewidth]{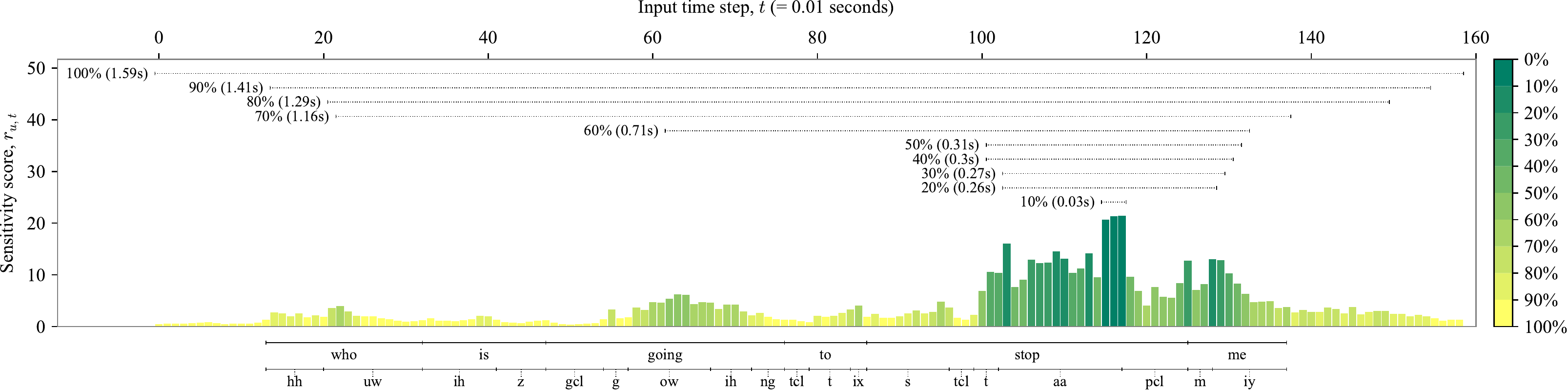}
  \caption{Sensitivity scores for the character ``p'' in the correctly predicted sentence ``who is  going to stop me'' by the CTC model trained on LibriSpeech. Hand-annotated word and phone alignments from the TIMIT dataset are shown in the bottom. The temporal spans corresponding to different levels of accumulated sensitivity are shown in the top. By averaging these across all non-blank character predictions in a test set, we obtain a measurement of the model's context sensitivity.} 
  \label{fig:sensitivity_example}
\end{figure*}
Connectionist temporal classification (CTC) \cite{graves2006connectionist} and attention-based encoder-decoder (AED) models \cite{chan2016listen, bahdanau2016end} are arguably the most popular choices for end-to-end automatic speech recognition (E2E ASR). 
However, it has been unclear if CTC and AED models process speech in qualitatively different ways.
The use of sentence-level context is important for human speech perception \cite{hutchinson1989influence}, but has not been studied for ASR.
Previous research has claimed that AED models learn a better implicit language model given enough training data \cite{battenberg2017exploring}.
Furthermore, comparisons of the two models have suggested that CTC models are inferior without the help of an external language model \cite{battenberg2017exploring, prabhavalkar2017comparison}, which leads to the hypothesis that CTC models are incapable of exploiting long temporal dependencies.

We study how the two E2E ASR models utilize temporal context. For this purpose, we consider first-order derivatives \cite{fu1993sensitivity, dimopoulos1995use} and the occlusion of input features \cite{zeiler2014visualizing}. While these methods have been frequently used to analyze natural language processing models \cite{arras2017explaining, li2015visualizing, arras2019evaluating, li2016understanding} their application to speech recognition has been limited \cite{krug2018introspection, bharadhwaj2018layer}.


We first highlight three general architectural differences between the two approaches and argue that these may enable AED models to utilize more temporal context than CTC models (section \ref{E2E}). Further, we define an intrinsic measure of context sensitivity based on the partial derivative of individual character predictions with respect to the input audio (section \ref{dba} and figure \ref{fig:sensitivity_example}). This allows us to analyze the sensitivity across the temporal dimension of the input for any E2E ASR model. Finally, we devise an experiment to directly compare model performance when context is constrained (section \ref{oba}). For this, we use hand-annotated word-alignments to accurately occlude temporal context. Our contributions are as follows:

\begin{enumerate}[leftmargin=*]

\item Through a derivative-based sensitivity analysis we show that the AED model is more context sensitive than the CTC model. Our ablation study attributes this difference to the attention-mechanism which closes the gap when applied to the CTC model. 


\item Although the AED model is more context sensitive than the CTC model without an attention-mechanism, we find that the two models perform similarly when contextual information is constrained by occluding surrounding words in the input audio. 



\item In contrast to previous comparisons, we show that the CTC model is highly competitive with the AED model without the help of an external language model. Using a deep and densely connected architecture, both models reach a new E2E state-of-the-art on the WSJ task.

\end{enumerate}



\clearpage

\section{End-to-End Speech recognition}
\label{E2E}

\subsection{Connectionist Temporal Classification Models}
\label{CTC}

Given a sequence of real valued input vectors $\mathbf{x}=(\mathbf{x}_{1}, ..., \mathbf{x}_{T})$, CTC models compute an output sequence $\mathbf{\hat y}=(\mathbf{\hat y}_{1}, ..., \mathbf{\hat y}_{U})$, where each $\mathbf{\hat y}_{u}$ is a categorical probability distribution over the target character set. Apart from the letters a-z, white-space and apostrophe ('), the character set also includes the special blank token (-). The input and output lengths, $T$ and $U$, are related by $U = \lceil \frac{T}{R} \rceil$ where $R$ is a constant reduction factor achieved by striding or stacking adjacent temporal representations. In this study, we never use an external language model. Instead, we rely on a simple greedy decoder $\beta(\cdot)$ that collapses repeated characters and removes blank tokens (e.g., $\textit{-c-aatt-} \mapsto \textit{cat}$). The $\beta(\cdot)$ function operates on the predicted alignment path $\mathbf{\hat q} = ({\hat q}_{1}, ..., {\hat q}_U)$ obtained by letting $\hat q_u = \argmax_{q}{\hat y}_{u, q}$.


This decoding mechanism results from the CTC loss function. The loss is computed by summing the probability of all alignment paths $\mathbf{q} = (q_{1}, ..., q_U)$ that translate to the target sequence $\mathbf{y}$. The probability of a single path is given by:
%
%
\begin{align}
    \textrm{P}(\mathbf{q} | \mathbf{x}) = \prod_{u=1}^{U}{\hat y}_{u, q_u}
\end{align}
\noindent Given the set of paths $\{\mathbf{q} \mid \beta(\mathbf{q}) = \mathbf{y}\} = \beta^{-1}(\mathbf{y})$ that translate to a given target transcript, the total probability is:
\begin{align}
    \textrm{P}(\mathbf{y} | \mathbf{x}) = \sum_{\mathbf{q} \in \beta^{-1}(\mathbf{y}) } \textrm{P}(\mathbf{q} | \mathbf{x})
\end{align}
\noindent The loss is simply $L(\mathbf{y}, \mathbf{\hat y}) = \ln (\textrm{P}(\mathbf{y} | \mathbf{x}))$ which can be computed efficiently with dynamic programming \cite{graves2006connectionist}. 

\begin{figure}[!b]
  \centering
  \includegraphics[width=\linewidth]{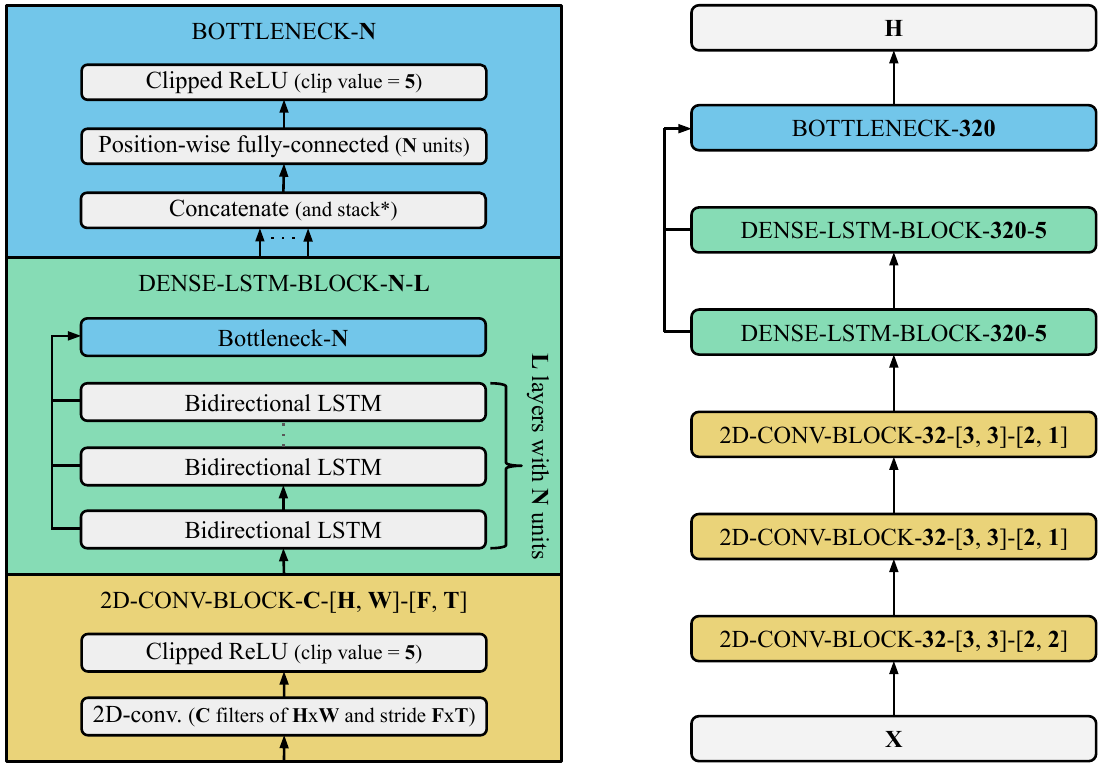}
  \caption{Default encoder architecture used for both CTC and AED models. * Only applied in the bottleneck layer of the first dense LSTM block for the AED model to achieve $R=4$.}
  \label{fig:architecture}
\end{figure}

\subsection{Attention-based Encoder-Decoder Models}
\label{AED}

AED models first encode the input $\mathbf{x}$ to a sequence of vectors $\mathbf{h} = (\mathbf{h}_{1}, ..., \mathbf{h}_{U}) = \textsc{encode}(\mathbf{x})$ which is passed to an autoregressive decoder function $\textsc{decode}(\cdot)$. We reuse $U$ to denote the length of $\mathbf{h}$ to emphasize that, as with CTC models, it is defined by a constant reduction factor $R$. However, AED models are typically robust to a high reduction factor ($R \leq 2^4$) compared to CTC models ($R \leq 2^1$) \cite{battenberg2017exploring}. Operating at a lower temporal resolution should make it easier for recurrent encoder layers (section \ref{arch}) to pass information across longer time spans.

Roughly speaking, we could write the decoder as $\mathbf{\hat y}_{k} = \textsc{decode}(\mathbf{h}, \mathbf{\hat y}_{k-1}, \mathbf{s}_{k-1}, \mathbf{a}_{k-1})$ where $\mathbf{\hat y}_{k}$ is a probability distribution over characters, $\mathbf{s}_{k}$ is the decoder state and $\mathbf{a}_{k}$ is the attention vector. Unlike CTC models, there are no repeated characters or blank tokens to interleave the final predictions. Thus, given the same output sequence, we have $K \leq U$. As with the encoder, lower temporal resolution between decoder steps could make it easier to pass information between predictions. 


Emphasizing more detail, we split the $\textsc{decode}(\cdot)$ function into the following sequence of computations:
\begin{align}
    \mathbf{s}_k &= \textsc{recurrent}(\mathbf{s}_{k-1}, [\Phi(\mathbf{\hat y}_{k-1}); \mathbf{a}_{k-1}]) \\
    \mathbf{a}_{k} &= \textsc{attend}(\mathbf{s}_k, \mathbf{h}) \\
    \mathbf{\hat y}_{k} &= \textsc{predict}(\mathbf{a}_{k})
\end{align}
\noindent Here $[\,\cdot\,;\,\cdot\,]$ denotes the concatenation of two vectors and $\Phi(\cdot)$ is a non-differentiable embedding lookup.\footnote{The lookup is not captured by the gradient-based sensitivity analysis presented in \ref{dba}.} The $\textsc{recurrent}(\cdot)$ function can take the form of any recurrent neural network architecture. We use a single LSTM \cite{hochreiter1997long} cell for all our experiments. The $\textsc{predict}(\cdot)$ function is a single fully-connected layer followed by the softmax function. The following steps deﬁne the $\textsc{attend}(\cdot)$ function:
\begin{align}
    e_{k,u} &= \mathbf{v}^\top \mathrm{tanh}(\mathbf{W}_{s}\mathbf{s}_k+\mathbf{W}_h\mathbf{h}_u) \\
    \alpha_{k,u} &= \frac{\mathrm{exp}(e_{k,u})}{\sum_{u'=1}^{U} \mathrm{exp}(e_{k,u'})} \\
    \mathbf{c}_k &= \sum_{u=1}^{U} \alpha_{k,u}\mathbf{h}_u \\
    \mathbf{a}_k &= \mathrm{tanh}(\mathbf{W}_{a}[\mathbf{c}_k;\mathbf{s}_k])
\end{align}
\noindent Where $\mathbf{v}$, $\mathbf{W}_{s}$, $\mathbf{W}_h$ and $\mathbf{W}_{a}$ are trainable parameters. The computation of the energy coefficient $e_{k,u}$ is taken from \cite{bahdanau2014neural}. Note that each energy coefficient, and thus each attention weight $\alpha_{k,u}$, is computed identically for all encoder representations $\mathbf{h}_u$. Unlike recurrent network connections, combining information from time steps far apart does not require propagating the information through a number of computations proportional to the distance between the time steps.

Thus, we have highlighted three components that could make AED models more context sensitive: (\RNum{1}) Encoder resolution, (\RNum{2}) decoder resolution and (\RNum{3}) the attention-mechanism.

\subsection{Encoder Architecture}
\label{arch}

Whereas the main contribution of the CTC framework is the loss function, the AED model relies on a more complex architecture that allows it to be trained with a simple cross-entropy loss. To see this, note that the CTC forward pass can be stated as a subset of the functions introduced in section \ref{AED}:
\begin{align}
    \mathbf{h} &= \textsc{encode}(\mathbf{x}) \\
    \mathbf{\hat y}_u &= \textsc{predict}(\mathbf{h}_{u})
\end{align}
\noindent As in previous work, we use convolutions followed by a sequence of bidirectional recurrent neural networks \cite{park2019specaugment, amodei2016deep, zhang2017very}. Our final encoder has 10 bidirectional LSTM layers with skip-connections inspired by \cite{huang2017densely}. The outputs of the forward and backward cells are summed after each LSTM layer. Default is $R=2$ for CTC and $R=4$ for AED. See figure \ref{fig:architecture}.



\section{Method}

We used two different approaches for analyzing temporal context utilization of the two E2E ASR models. The derivative-based sensitivity analysis (\ref{dba}) can be used to compare a set of models on any dataset. However, as we will see, there is no guarantee that the differences found with this approach translate to better performance. The occlusion-based analysis (\ref{oba}) allows us to evaluate how the models respond when we remove temporal context. This measure is easy to interpret and can be used to directly asses the importance of temporal context, but requires hand-annotated word-alignments which are rarely available in publicly available datasets.

\begin{table}[t]
\begin{center}
\begin{tabular}{ l c c c r} 
\toprule
\textbf{Model} & \textbf{clean} & \textbf{other} & \textbf{type} & \textbf{params} \\
\midrule
Li et al., 2019 \cite{li2019jasper} & 3.86 &  11.95 & CTC & 333 M \\
Kim et al., 2019 \cite{kim2019improved} & 3.66 & 12.39 & AED & \texttildelow 320 M \\
Park et al., 2019 \cite{park2019specaugment} & 2.80 & 6.80 & AED & \texttildelow 280 M \\
\midrule
Irie et al., 2019 \cite{irie2019choice} \\
\footnotesize\hspace{3mm}Small - Grapheme & 7.9 & 21.3 & AED & 7 M \\
\footnotesize\hspace{3mm}Small - Word-piece & 6.1 & 16.4 & AED & 20 M \\
\footnotesize\hspace{3mm}Medium - Grapheme & 5.6 & 15.8 & AED & 35 M \\
\footnotesize\hspace{3mm}Medium - Word-piece & 5.0 & 14.1 & AED & 60 M \\
\midrule
\textit{Our work:} \\
Deep LSTM & 5.13 & 16.03 & CTC & 17.7 M \\
Deep LSTM & 5.45 & 17.05 & AED & 19.8 M \\
\bottomrule
\end{tabular}
\end{center}
\caption{Word error rates on the clean and other test sets of LibriSpeech. None of the above use an external language model.}
\label{tab:libri}
\end{table}

\subsection{Derivative-based Sensitivity Analysis}
\label{dba}

We define a sequence of sensitivity scores $\mathbf{r}_{k}$ ($\mathbf{r}_{u}$ for CTC models) across the temporal dimension of the input space for each predicted character. Let $F$ be number of spectral input features and $Q$ the size of the output character set:
\begin{align}
    r_{k, t} &= \sum^Q_{q=1}\sum^F_{f=1}\left\lvert \frac{\partial \hat y_{k,q}}{\partial x_{t,f}} \right\rvert
\end{align}
\noindent An example is shown in figure \ref{fig:sensitivity_example}. Our goal is to measure the dispersion of these scores across the input time steps. We do so by summing the scores from largest to smallest and measure the temporal span of the scores accumulated for a certain percentage of the total sensitivity. For example, the set of scores needed to account for $10\%$ of the total sensitivity may be $\{r_{k,3}, r_{k,7}, r_{k,8}, r_{k,10}\}$. The temporal span would then be $10-3=7$ time steps corresponding to $0.07$ seconds. We take the mean of this span for a fixed percentage across all character predictions in a given data set to summarize the temporal context sensitivity of a model. This allows us to evaluate how the sensitivity disperses as we increase the accumulated percentage. A higher dispersion of sensitivity scores equals a higher context sensitivity. The derivative-based measure considers a linearization of the models and, thus, does not capture non-linear effects.


\begin{figure*}[t]
  \centering
  \includegraphics[width=\linewidth]{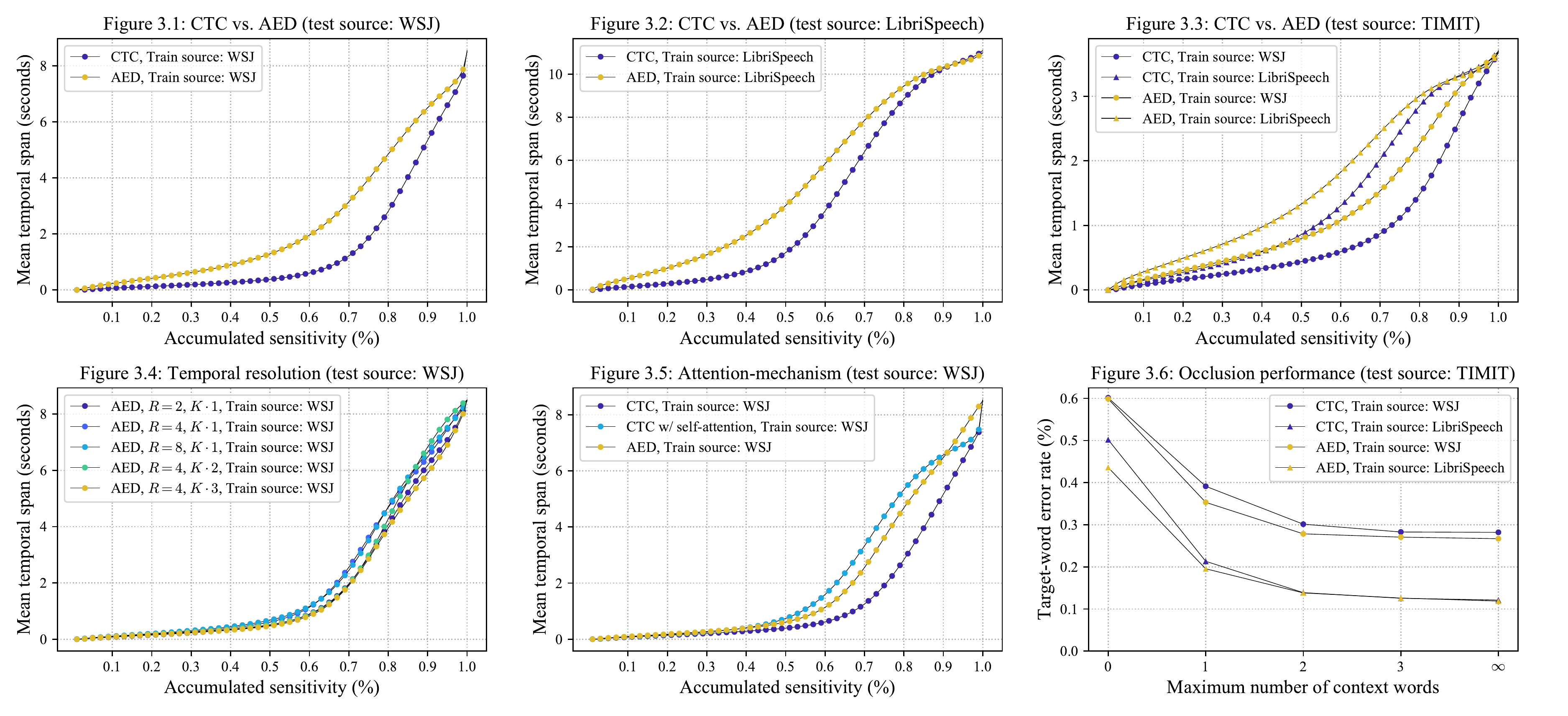}
  \caption{
            Sensitivity analysis (figures 3.1-3.5) and occlusion-based analysis (figure 3.6). See corresponding subsections.
  }
  \label{fig:model_and_source}
\end{figure*}

\begin{table}[t]
\begin{center}
\begin{tabular}{ l c c } 

\toprule
\textbf{Model} & \textbf{eval92} & \textbf{type}\\
\midrule
Chorowski \& Jaitly, 2016 \cite{chorowski2016towards} & 10.60 & AED\\
Zhang et al., 2017 \cite{zhang2017very} & 10.53 & AED \\
Chan et al., 2016 \cite{chan2016latent} & 9.6 & AED \\
Sabour et al., 2018 \cite{sabour2018optimal} & 9.3 & AED \\
\midrule
Our work: \\
Deep LSTM & 9.25 & CTC \\
Deep LSTM & 9.25 & AED \\
\bottomrule

\end{tabular}
\end{center}
\caption{Word error rates on the eval92 test set of WSJ. None of the above use an external language model.}
\label{tab:wsj}
\end{table}

\subsection{Occlusion-based Analysis}
\label{oba}
To directly test the dependence on contextual information, we use hand-annotated word-alignments to systematically occlude context. Given a word $w_t$, we test how well a model recognizes the word given different levels of context. That is, we crop out the audio segment corresponding to $w_{t-C}, ..., w_{t+C}$ where $C$ is the maximum number of context words visible on each side.\footnote{We also add the silence from the start and end of the original sentence to the audio segment as it improves model performance.} If the target word $w_t$ is in the predicted sequence, we accept the hypothesis. To avoid ambiguous situations where the target word is identical to one of the $2C$ context words, we only make use of sentences that consist of a sequence of unique words. 

\section{Experiments}
\label{experiments}
\subsection{Data and Training}
We trained the models on the Wall Street Journal CSR corpus (WSJ) \cite{paul1992design} and the LibriSpeech ASR corpus \cite{panayotov2015librispeech}. WSJ contains approximately 81 hours of read newspaper articles and LibriSpeech contains 960 hours of audio book samples. We used 80-dimensional log-mel spectrograms as input. The models were trained for 600 epochs on WSJ and 120 epochs on LibriSpeech. We used Adam \cite{kingma2014adam} with a fixed learning rate of $3\cdot 10^{-4}$ for the first 100 epochs on WSJ and 20 epochs on LibriSpeech, before annealing it to $1/6$ of its original size. We used dropout after each convolutional block \cite{tompson2015efficient} and each bidirectional LSTM layer \cite{kingma2015variational}. The dropout rate was set to 0.10 for models trained on LibriSpeech and 0.40 for WSJ. Similar to \cite{schneider2019wav2vec}, we constructed batches of similar length samples, such that one batch consisted of up to 320 seconds of audio and contained a variable number of samples. For the AED model, we used teacher-forcing with a 10\% sampling rate.

For the occlusion-based analysis, we considered the hand-annotated word-alignments from the TIMIT dataset \cite{garofolo1993timit}. We  excluded all sentences repeated by multiple speakers in order to avoid biasing the results towards certain sentence constructions (i.e., we only use the SI-files of the TIMIT dataset).

\subsection{ASR results}
\label{asr-performance}

We compare the default configuration of our CTC and AED models trained on WSJ and LibriSpeech to other notable E2E ASR models in table \ref{tab:libri} and \ref{tab:wsj}. Both the CTC and AED model compare favorably to more sophisticated approaches on WSJ. On LibriSpeech, our models do not perform as well as larger models, but are still on par with models of comparable size from \cite{irie2019choice} which is the same model as in \cite{park2019specaugment} at smaller scale. The slightly worse performance of the AED model on LibriSpeech can be attributed to longer sentences which have a tendency to destabilize training. Similar issues have been reported in prior work \cite{chan2016listen, battenberg2017exploring}.

\subsection{CTC vs.~AED}
\label{CTCvsAED}
As hypothesized, figure 3.1 and 3.2 reveal that our AED models utilized a larger temporal context than the CTC models based on the sensitivity scores. The trend was consistent across all levels of accumulated sensitivity scores. In figure 3.3, we see the same pattern when evaluated on the TIMIT dataset which will be used for the occlusion-based analysis.

\subsection{Temporal resolution}

We trained the AED model with three different temporal encoder resolutions $R = 2, 4, 8$ on the WSJ dataset. $R$ was configured by increasing stride in each of the three convolutional layers. As seen in figure 3.4, encoder resolution had no impact on context sensitivity.

To test decoder resolution, we interleaved the target transcript with one or two redundant blank tokens to effectively increase the target length to $K\cdot2$ or $K\cdot3$. Figure 3.4 shows that decoder resolution had no impact on context sensitivity.

\subsection{Attention-mechanism}

To test how the attention-mechanism affects context sensitivity, we incorporated the $\textsc{attend}(\cdot)$ function in the CTC architecture. Instead of passing $\mathbf{h}_u$ directly to $\textsc{predict}(\cdot)$, we first applied self-attention:
\begin{align}
    \mathbf{\hat y}_u &= \textsc{predict}(\textsc{attend}(\mathbf{h}_u, \mathbf{h}))
\end{align}
\noindent We trained this model on the WSJ dataset and compared it to the AED model and the CTC model without attention in figure 3.5. The attention-mechanism closed the gap in context sensitivity between the two models. Thus, the difference found in sections \ref{CTCvsAED} is likely a result of this architectural component that can be easily incorporated in a CTC model. However, a large $U$ results in high memory consumption. Therefore, we used a smaller model where the two dense LSTM blocks are replaced by three LSTM layers with 128 units for the experiments shown in figures 3.4 and 3.5.

\subsection{Occlusion performance}
Figure 3.6 shows how model performance is affected under different context constraints. We see that both the CTC and AED model suffered severely when contextual information was completely removed. The models came close to optimal performance when approximately three words were allowed on each side of the target word. Thus, temporal context is an important factor for both models. This result aligns well with the common n-gram size (3-4) when decoding with the help of a statistical language model \cite{chorowski2016towards, zeghidour2018fully, luscher2019rwth}.

Based on the results in section \ref{CTCvsAED}, we would expect that the AED models rely more on the temporal context than the CTC model. However, we do not see such a trend in figure 3.6. Indeed, there was no pronounced or consistent difference between the two models regardless of training source. This result implies that the architectural differences between the AED and CTC models do not necessarily translate to a performance difference. It may be that the AED model included more evidence from context than the CTC model, but the results in figure 3.6 indicate that this did not add any additional value in terms of lowering word error rate.




\section{Conclusions}
We show that AED models are generally more context sensitive than CTC models and that this difference is largely explained by the attention-mechanism of AED models. Adding a self-attention layer to the CTC model bridges the gap between the models. Analyzing performance by constraining temporal context, we also find that the initial difference between the two models is not crucial in terms of word error rate performance, although both models rely heavily on context for optimal performance. Our experiments on WSJ and LibriSpeech show that CTC models are capable of delivering state-of-the-art results on par with AED models without an external language model. 
Because of its simplicity and more stable training, CTC is our preferred E2E ASR framework.
\bibliographystyle{IEEEtran}

\bibliography{mybib}


\end{document}